\documentclass[preprint,12pt]{elsarticle}

\usepackage{amssymb}

\usepackage{lineno}

\usepackage{color}
\usepackage[dvipsnames]{xcolor}
\usepackage{algorithmic,algorithm}
\usepackage{url}
\usepackage{multirow}
\usepackage{amsmath}

\journal{Information Sciences}

\begin{document}

\begin{frontmatter}

\title{Interactive Multi-Objective Evolutionary Optimization of Software Architectures}


\author{Aurora Ram\'irez\corref{x}}\ead{aramirez@uco.es}
\author{Jos\'e~Ra\'ul~Romero\corref{corauthor}}\ead{jrromero@uco.es}
\cortext[corauthor]{Corresponding author. Tel. +34 957 21 26 60}
\author{Sebasti\'an Ventura\corref{x}}\ead{sventura@uco.es}

\address{Department
of Computer Science and Numerical Analysis, University of C\'ordoba, 
14071, C\'ordoba, Spain}

\begin{abstract}
While working on a software specification, designers usually need to evaluate different architectural alternatives to be sure that quality criteria are met. Even when these quality aspects could be expressed in terms of multiple software metrics, other qualitative factors cannot be numerically measured, but they are extracted from the engineer’s know-how and prior experiences. In fact, detecting not only strong but also weak points in the different solutions seems to fit better with the way humans make their decisions. Putting the human in the loop brings new challenges to the search-based software engineering field, especially for those human-centered activities within the early analysis phase. This paper explores how the interactive evolutionary computation can serve as a basis for integrating the human’s judgment into the search process. An interactive approach is proposed to discover software architectures, in which both quantitative and qualitative criteria are applied to guide a multi-objective evolutionary algorithm. The obtained feedback is incorporated into the fitness function using architectural preferences allowing the algorithm to discern between promising and poor solutions. Experimentation with real users has revealed that the proposed interaction mechanism can effectively guide the search towards those regions of the search space that are of real interest to the expert.
\end{abstract}


\end{frontmatter}


\section{Introduction}\label{sec:intro}

Making decisions is an intrinsic aspect of any software design task, since engineers have to choose the best design alternative among all the possibilities on the basis of both functional and non-functional requirements. During the architectural analysis, abstract artifacts need to be precisely identified and specified in order to efficiently guide the development, evolution and deployment of the overall system. Considering such an early stage, architectural decisions become even more challenging due to the lack of knowledge about the system but, at the same time, they are crucial to fulfill the many quality criteria imposed~\cite{Falessi2011}.

Artificial intelligence techniques and, more specifically, metaheuristics, can support software engineers in their decision processes by providing them with effective methods to explore a great deal of software designs, each one determined by a different trade-off among the required quality aspects. Such a scenario can be viewed as one of the goals of the search-based software engineering (SBSE) field~\cite{Harman2012}, in which optimization techniques are applied to the resolution of software engineering (SE) tasks conveniently reformulated as search problems. However, solving human-centered activities in a fully automated way seems to be unrealistic, especially for those related to the analysis phase. Certainly, trying to capture the richness of human knowledge only by means of software metrics still represents an unresolved matter to the SE community~\cite{Simons2015}. Hence, most of the evaluation methods proposed at the architectural level strongly rely on the expert's judgment~\cite{Dobrica2002}, making extremely difficult to precisely formulate a quantitative fitness function.

Given the relevance of the software architect for the design process, search-based approaches should benefit from his/her knowledge and expertise in order to address the optimization problem in the same way s/he would do it. Interactive optimization~\cite{Meignan2015} constitutes a compelling paradigm here. It allows the human to actively participate in such a way that the expert's opinion can influence both the problem formulation and the search process, allowing the adaptation of the interaction mechanism to the specific requirements of the application domain.
 
Due to the many various aspects involved with architectural analysis, related optimization problems often require the definition of multiple conflicting objectives. In this sense, the integration of interactive approaches into multi-objective evolutionary algorithms (MOEAs)~\cite{Coello2007} needs further considerations. Since a MOEA requires the presence of a \textit{decision maker} (DM) in order to choose the final solution among the set of alternatives returned, a logical step would be to allow the DM to dynamically express his/her preferences during the search process~\cite{Miettinen2008}.

Even when maintaining a multi-objective perspective for architectural optimization is clearly necessary, conceiving an interaction mechanism only founded on expressing opinions about the objective space seems to be insufficient. In addition, comparing several architectural models becomes a hard task due to the information overload. However, engineers would feel more confident when they strictly evaluate qualitative aspects of the automatically generated architectural solutions~\cite{Santhanam2016}. This approach permits them to extend the scope of the feedback provided to the algorithm, as well as to adapt the sort of requested opinion as the search elapses. For instance, delivering both positive and negative judgments, which perfectly matches with the human way of acting, can assist the algorithm to discern between interesting and poor solutions with respect to the expert's understanding.

In this context, this paper proposes an interactive evolutionary approach to address the so-called discovery of component-based software architectures~\cite{Ramirez2015a}. In this problem several software metrics based on maintainability are considered for the evaluation of structural aspects of the components and interfaces that constitute the early software specification. Nevertheless, assessing the adequacy of these highly-abstract software artifacts should also rely on the engineer's feedback. With these factors in mind, the following two research questions (RQ) were stated:

\emph{RQ1: How can the qualitative judgment of the engineer be integrated into the evolutionary discovery of software architectures?} The proposed interactive approach should consider the multi-objective nature of the optimization problem and define an appropriate evaluation mechanism, in which both qualitative and quantitative evaluation criteria could be put together.

\emph{RQ2: Does putting the human in the loop involve a significant improvement compared with not considering him/her along the optimization process?} The interactive system should strike a balance between the evolutionary performance, measured by usual quality indicators, and the practical incentive for the software engineer in terms of a reasonable number of high-quality solutions satisfying his/her preferences. Such an analysis requires conducting an empirical study with a substantial number of participants, where aspects like usefulness and intuitiveness should be also evaluated.

A main contribution of this work is the combination of qualitative and quantitative evaluation criteria. On the one hand, an interactive system manages design decisions entered by the engineer to either intensify the search towards specific regions of the search space or, on the contrary, escape from those that do not meet his/her expectations. More specifically, each design decision is mapped into a function, named \emph{architectural preference}, that reinforces the fitness value of solutions satisfying the corresponding qualitative characteristic. On the other hand, the quantitative evaluation in terms of software metrics is kept as a means for achieving promising candidate solutions from a multi-objective perspective. Additionally, it serves to control the inherent uncertainty that arise when dealing with human reasoning, such as fatigue and inconsistency~\cite{Parmee2001}. Reports obtained from real user experiences show that the interactive algorithm here presented is able to adapt the search as new design decisions are made.

The rest of the paper is structured as follows. Section~\ref{sec:background} briefly introduces software architecture optimization methods, as well as the concepts and terminology related to the interactive evolutionary computation (IEC) and its application to SBSE. Section~\ref{sec:problem} describes the optimization problem under study, while the interactive evolutionary approach for discovering software architectures is detailed in Section~\ref{sec:algorithm}. Next, Section~\ref{sec:experimentation} presents the 
empirical method and experimental framework. Experiments assessing both the evolutionary performance and the applicability of the approach are presented in Section~\ref{sec:results}. Finally, threats to validity are discussed in Section~\ref{sec:threats}, and Section~\ref{sec:conclusion} concludes.

\section{Background}\label{sec:background}

This section explains the basis of how search techniques have been previously applied to address architectural design problems, describing some non-interactive optimization approaches. However, interactive optimization systems propose a completely different perspective to face optimization problems, involving the human in the search process. This fact clearly influences their design and implementation, briefly discussed in this section too. Next, some related work focused on the use of interactive approaches in SBSE is introduced.

\subsection{Software architecture optimization}\label{subsec:architectures}

During early analysis, the conception of a software architecture satisfying both the functional and non-functional requirements constitutes an important activity. Apart from describing the abstract structure of the system, an architecture exposes the design principles that should guide its subsequent development and evolution~\cite{Garlan2000}. Similarly, architectural models represent essential artifacts when addressing other activities of the software life cycle, such as resource allocation during deployment or reconstruction as part of maintenance and migration~\cite{Ducasse2009}. Carrying out these tasks as optimization problems is the idea behind the application of software architecture optimization methods~\cite{Aleti2013}. 

Optimization methods like metaheuristics can be used to arrange elements of an architectural specification or semi-automatically derive new models. This is done according to predefined quality attributes and other existing constraints. The advantage of applying these methods lies on their high capacity to explore a wide set of design alternatives, only requiring minor adaptation in order to properly manage the problem-specific decision variables. Nevertheless, the abstract and cross-cutting nature of software architectures need to be thoroughly observed to satisfactorily support the decision-making process~\cite{Falessi2011}, in which it may influence other factors like human intuition, conflicting goals or the uncertainty inherent to this early stage.

Software architecture optimization has recently emerged as an upward trend in SBSE, providing the necessary support to software engineers when dealing with complex design scenarios. Recent advances reveal that multi-objective evolutionary algorithms can be effectively applied to enhance architectural artifacts at different stages of the design process. For instance, NSGA-II has served to assist engineers in the production of architectural documentation~\cite{Diaz-Pace2016}. A hybrid approach considering analytical optimization and a variant of NSGA-II was also presented to cope with the selection and allocation of software components during deployment~\cite{Koziolek2013}. Similarly, reconfiguration after deployment was defined as a 5-objective optimization problem to be solved by a specific genetic algorithm~\cite{Vescan2016}.

\subsection{Interactive optimization}
\label{subsec:iec}

Interactive optimization encompasses all those search methods in which a human explicitly takes part in the search~\cite{Meignan2015}. The need for involving the human within the process can be motivated by many different factors, such as the inability to capture complex features around the problem formulation or the lack of an appropriate quantitative fitness function. This latter issue constitutes a major concern when attempting to solve creative tasks by means of evolutionary computation (EC), so it is not surprising that initial efforts were mostly focused on leaving the responsibility for evaluating candidate solutions to humans~\cite{Takagi2001}. In these cases, showing a subset of the population and then interpolating the fitness for the rest of individuals is a common strategy to reduce the cognitive burden.

When addressing multi-objective problems (MOPs), DMs are expected to establish the desired trade-off among objectives either at the beginning of, during or after the search. To this end, several methods has been proposed, including the negotiation of the importance of each objective and the definition of reference points~\cite{Miettinen2008}. The gathered information would be used to redirect the search towards certain regions of the Pareto front (PF) or even to learn from the DM's preferences~\cite{Branke2015}. Although these mechanisms have been already integrated into some existing MOEAs, other algorithms specifically conceived to deal with MOPs from an interactive perspective can be also found in the literature. A representative approach is iTDEA (\emph{interactive territory defining evolutionary algorithm})~\cite{Koksalan2010}, which progressively delimits preferred regions of the PF around the most interesting solutions. More specifically, some solutions are presented to the DM at certain moments of the search process in order to choose the best. According to the feedback obtained, iTDEA updates the size of the territory associated to similar solutions, which determines the permitted distance between the individuals stored in an external archive.

Notice that interactive multi-objective optimization usually restricts the human's decisions to the objective space. However, humans may feel uncertainty about their own opinion when the definition of the objective functions is not easily understandable. Therefore, contributing with opinions on qualitative criteria would fit better in those cases where the interest lies on aspects of the solution to be evaluated from the expert's point of view. This approach is considered in~\cite{Brintrup2008}, where subjective criteria are considered to define an objective function that is computed together with another function determined by quantitative criteria. In each iteration, the expert rates each solution using a 0-9 scale and can perform additional actions such as altering solutions. Other ways of integrating qualitative information are based on fuzzy modeling of user's preferences and rule-based systems~\cite{Takagi2001}.

\subsection{Interactive approaches in SBSE}\label{subsec:isbse}

Software engineering seems to represent a natural scenario for interactive optimization, since most of its activities are traditionally carried out by human beings. In fact, real experiences reported by recent works confirm the suitability of these kind of methods~\cite{Marculescu2015,Simons2014}, showing the interest of the SE community in the development of decision support systems under the SBSE paradigm. A recent review of the state-of-the-art~\cite{Ramirez2018} also highlights that evaluation mechanisms based on qualitative preferences, whether they have been freely expressed by the expert or selected from a set of options, are preferred over the direct assignment of fitness values.

Software engineers' abilities and know-how are specially important when tackling analysis and design tasks. Therefore, it makes sense that the majority of the interactive approaches proposed in SBSE belong to the so-called search-based software design subfield~\cite{Raiha2010}. For instance, interactive conceptual object-oriented design has been successfully addressed using both EC~\cite{Simons2012} and ant colony optimization (ACO)~\cite{Simons2014}. Here, the authors examine the influence of aesthetic criteria, defined in terms of elegance metrics, when class diagrams are derived from use cases. This work was then extended to allow the designer to freeze parts of the solution, demonstrating the effectiveness of ACO to obtain high-quality solutions after a small number of iterations.

Architecture synthesis~\cite{Vathsavayi2013} and software refactoring~\cite{Mkaouer2014} are other examples of design problems addressed using interactive approaches. On the one hand, the interaction mechanism employed in~\cite{Vathsavayi2013} allows the architect to freeze classes and design patterns in order to compose the low-level architecture of a software system. Using class diagrams to represent the solutions, they are only evaluated in terms of quantitative software metrics. On the other hand, a multi-objective approach is used in~\cite{Mkaouer2014} with the aim of improving code quality. Firstly, NSGA-II is responsible for approximating the whole PF. Then, an interactive mechanism assists the engineer in identifying the most interesting refactoring sequences, as they represent the input information required by a local search procedure.

\section{Problem fundamentals}\label{sec:problem}

This section describes in detail the search problem for the evolutionary discovery of software architectures. Adopted from our previous work~\cite{Ramirez2015a}, where an initial non-interactive evolutionary solution was proposed, the encoding of candidate solutions and the metrics for their quantitative evaluation are also explained.


\subsection{The search problem}\label{subsec:problem}

Understanding the original architecture of a system as it evolves becomes a complex task if the corresponding analysis information is not properly generated and maintained. There are also situations in which the software engineer just needs to specify software artifacts at a higher level of abstraction as an important step prior to the addition of new functionality or the migration of the system. Software components represent abstract units of construction providing well-defined services that can be accessed through their interfaces~\cite{Szyperski2002}. Promoting reusability is the ultimate goal of organizing the system structure this way.

In this context, the discovery of component-based software architectures consists in identifying the high-level structure of a software system, in terms of its components and interfaces, from a previous analysis model represented by a UML~2 class diagram~\cite{Ramirez2015a}. More specifically, the discovery process can be defined according to the following rules:

\begin{itemize}
	\item A component is derived from a cohesive group of classes, all of them working together in order to implement its behavior.

	\item Directed relationships among classes belonging to different components serve to identify interfaces, since they represent the provision or need of functionalities. Public methods within classes are used to determine the interface operations. When two or more classes are involved in an interaction between a pair of components, all their public methods would specify a unique interface. Notice that the previous model described in~\cite{Ramirez2015a} established that each pair of related classes represents a candidate interface. In contrast, the approach here presented has been improved to increase flexibility in such a way that, if the same class is required by other components, its public methods are properly separated following reusability criteria.
    
	\item A connector represents the linking between different components in terms of their matching interfaces. As a consequence of how interfaces are now derived, a connector could link a provided interface to several required interfaces.
\end{itemize}

Following these rules, the search algorithm is able to explore the decision space looking for the optimal allocation of classes and their relationships into components and interfaces. Even so, the abilities and know-how of the software engineer play a key role when envisioning the main functionalities of large software systems and their mutual interactions. Thus, the discovery of software architectures is still a human-centered and iterative task. 

\subsection{Encoding and initialization}\label{subsec:init}

Each candidate solution represents a complete component-based software architecture, whose phenotypic expression corresponds to its representation as a UML~2 component diagram. As for the search, each solution is encoded using a tree structure, whose hierarchical composition perfectly reflects how an artifact, e.g. a component, is comprised of elements of a lower level, e.g. classes and interfaces.

All the encoded solutions should satisfy a number of constraints in order to represent feasible architectural models, as defined next: 

\begin{itemize}
	\item Each class must be only located into one component. Components cannot be empty.
	\item Components should define at least one interface, either required or provided.
	\item A pair of components could not provide services to each other, so that they would be mutually dependent.
\end{itemize}

The population is initialized by arbitrarily distributing classes from the input diagram into a random number of components within the range set by the software engineer. The process controls that classes are not replicated and no empty component is returned. However, constraints regarding inadmissible interactions among components are not checked. These unfeasible individuals will be penalized in terms of their fitness value, making them progressively disappear.

\subsection{Software metrics for quantitative evaluation}\label{subsec:metrics}

\begin{table*}[!t]
\renewcommand{\arraystretch}{1.3}
\begin{center}
\caption{Software metrics to evaluate component-based architectures}
\begin{tabular}{|c|}
\hline
$ICD=1/n\cdot\sum_{i=1}^{n}{\left((\#cl_{t}-\#cl_i)/\#cl_{t}\right) \cdot \left(CI_i^{in}/(CI_i^{in} + CI_i^{out}) \right)}$ \\
\hline
$ERP=\sum_{i=1}^{n}{\sum_{j=i+1}^{n}{(w_{as}\cdot \#as_{ij} + w_{ag}\cdot \#ag_{ij}+ w_{co}\cdot \#co_{ij} + w_{ge} \cdot \#ge_{ij})}}$ \\
\hline
$GCR=\#cgroups/n$\\
\hline
\end{tabular}
\label{tab:metrics}
\end{center}
\end{table*}

Dealing with several quality attributes like modularity or reusability in the same design is a usual situation for the architect. Hence, the discovery of software architectures can be addressed from a multi-objective perspective, where quality metrics measuring such attributes need to be simultaneously optimized. Table~\ref{tab:metrics} shows the three metrics, i.e. objectives, used in this work:

\begin{itemize}
	\item \textit{Intra-modular Coupling Density} (ICD) looks for a trade-off between coupling and cohesion. For each component $i$ it calculates the ratio between internal ($CI_i^{in}$) and external ($CI_i^{out}$) relations, also considering the number of inner classes ($\#cl_i$). $\#cl_t$ stands for the total number of classes in the full model, and $n$ is the number of components. This metric has to be maximized, and varies in the range [0,1].
    
	\item \textit{External Relations Penalty} (ERP) counts the number of relationships between classes of different components, $i$ and $j$, that cannot be declared with a well-defined interface. This situation mostly occurs when the navigability of associations ($as$), aggregations ($ag$) or compositions ($co$) is not explicitly defined in the source analysis model. Generalization relationships ($ge$) between classes located in different components, e.g. representing a data abstraction, may also turn into external dependencies. The software architect can specify how detrimental a dependency caused by each kind of relationship is for the resulting architecture in terms of their respective weights ($w_x$). ERP should be minimized, 0 being the optimum.

	\item \textit{Groups/Components Ratio} (GCR) defines the ratio between the number of groups of interconnected classes ($\#cgroups$) and the number of components in the entire architecture ($n$). A well-defined component is expected to contain a unique group of classes, so the optimal value is 1. This metric should be minimized.
\end{itemize}

\section{Interactive model for the evolutionary discovery of software architectures}\label{sec:algorithm}

As stated in \emph{RQ1}, considering the feedback from the software engineer into the discovery of software architectures would permit adding new valuable qualitative information to the search process. With this aim, this section provides an overview of the proposed IEC model. Then, a more detailed description of its essential elements is presented, including the solution evaluation method based on both quantitative and qualitative criteria, the mechanisms enabling the management and transformation of solutions, and how human interaction is conducted.

\subsection{Overview of the approach}\label{subsec:overview}

\begin{figure}[!t]
	\centering
	\includegraphics[width=0.7\textwidth]{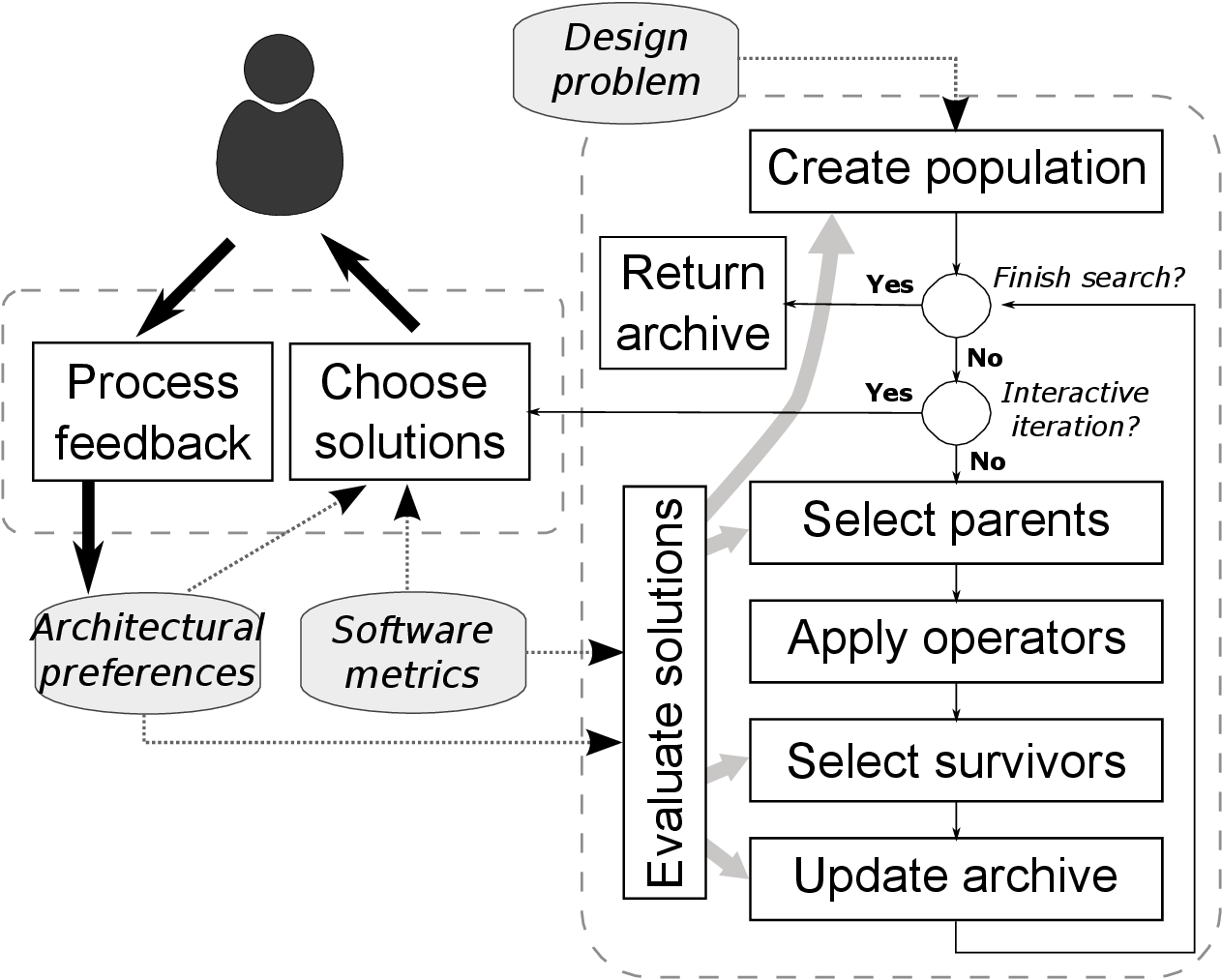}
	\caption{Proposed interactive evolutionary model.}
	\label{fig:overview}
\end{figure}

Fig.~\ref{fig:overview} shows the proposed evolutionary model (henceforth named iMOEA), which is composed of two main elements: the algorithm conducting the automatic search of architectural models, and the interaction module that coordinates the communication between the algorithm and the software engineer. More specifically, the multi-objective evolutionary algorithm here proposed is based on a steady-state scheme, and makes use of a sophisticated diversity preservation technique similar to hyperboxes. Both aspects are relevant according to our previous findings~\cite{Ramirez2015b}, since they provided a more appropriate convergence and a control mechanism to reduce the archive size, respectively. In addition, the algorithm defines a specific evaluation method that combines quantitative (software metrics) and qualitative (architectural preferences) criteria. The resulting fitness function is then used as one criterion to compare solutions along the search process.

Firstly, the algorithm initializes the population according to the procedure detailed in Section~\ref{subsec:init}. Notice that, at that time, solutions can be evaluated by only meeting quantitative criteria, i.e. the software metrics serving as objectives (see Section~\ref{subsec:metrics}). An initial archive is also created from the set of non-dominated solutions. Then, the evolutionary search follows the usual execution flow: parent selection, genetic operators, replacement and archive update. The search continues until a stopping condition is met.

At certain moments of the evolution, the algorithm momentarily stops the search to obtain feedback from the software engineer. During the interaction, the algorithm selects a subset of the population to be evaluated in terms of qualitative criteria. Architectural preferences are then defined in such a way that the algorithm will be able to numerically determine to what extent each candidate solution satisfies these criteria. The engineer can also perform additional \emph{actions}, such as freezing some specific elements of the architectural model under evaluation or even stopping the search. Notice that his/her choice might influence the course of the evolution in steps like the generation of offspring. After the first interaction, the evaluation phase begins to consider both quantitative and qualitative criteria, incorporating the expert's perspective in the optimization process.

\subsection{Fitness function: putting together human decisions and software
metrics}\label{subsec:fitness}

The lack of consistency and the cognitive burden, which are inherent in any interactive system, become even more critical when dealing with software architectures due to the presence of highly abstract artifacts. All this, combined with the huge amount of solutions that an algorithm can generate in one single execution, implies that relying only on the engineer's judgment to assess the quality of the solutions would be impractical. Thus, an effective evaluation mechanism requires striking the right balance between objective and subjective criteria. Software design metrics have proven to be effective in identifying the overall functional blocks of the architecture~\cite{Ramirez2015a}. As the design progresses and a more fine-tuned design is required, the participation of the expert becomes more relevant. Therefore, a reward/penalization approach~\cite{Ramirez2018} is adopted to capture the engineer's expectations. Given that both qualitative and quantitative assessments should be combined and computed, the feedback provided by the expert should be then mapped into numerical preference functions. With this aim, Eq.~\ref{eq:fitness} defines the fitness function of a solution $s$ as a weighted sum of two terms, $f_{obj}$ and $f_{sub}$. Weights $w_{obj}$ and $w_{sub}$ are considered in order to let the engineer control the relative importance of the objective and subjective evaluation, respectively. 
This function varies in the range [0,1] and should be minimized. Next, each component of the fitness function is explained in detail.

\begin{equation}
fitness(s)=w_{obj}\cdot f_{obj}(s) + w_{sub} \cdot f_{sub}(s)
\label{eq:fitness}
\end{equation}

\subsubsection{Objective evaluation: software metrics}\label{subsubsec:maximin}

The \emph{objective component}, $f_{obj}$, requires the definition of a set of software metrics, each one representing a conflicting objective. Without limiting the generality, $f_{obj}$ considers that all of them should be minimized and vary in the range [0,1]. Consequently ERP and GCR have been scaled accordingly. 
Their theoretical upper limit will depend on the number of relationships and the number of classes contained by the source analysis model, respectively. In addition, ICD values need to be inverted.

Given that obtaining a weighted sum of the objective values would target the search towards a unique point in the PF, a different kind of aggregation function is required here in order to transform the metric information into a single value. With this aim, $f_{obj}$ uses the \emph{maximin} function~\cite{Balling2001}. For a given set of objectives $k \in [1,K]$ this function returns a value in the range [-1,1], reporting on both the dominance and the diversity of a solution with respect to a reference set $Z$, e.g. the whole population. On the one hand, the sign of the result serves to distinguish between non-dominated ($<0$), weakly-dominated ($=0$) and dominated solutions ($>0$). On the other hand, the specific value gives an idea of the proximity between non-dominated solutions, values close to -1 being preferred. For a dominated solution, the result represents its distance to the PF, values close to 0 meaning proximity. Both properties serve to precisely qualify a candidate solution from a multi-objective perspective and, at the same time, the obtained values can be easily interpreted. Eq.~\ref{eq:fmetrics} shows the formulation of the \emph{maximin} function, properly adapted to return a value in the range [0,1], as required by Eq.~\ref{eq:fitness}. $f_k^s$ represents the value of the objective $k$ for the solution under evaluation ($s$), whereas $f_k^z$ represents the same value but for a solution, $z$, belonging to the reference set ($Z$). Notice that the \emph{maximin} function inspects all the search directions in order to find which one allows the solution $s$ to be far from being dominated.

\begin{equation}
f_{obj}(s)=(1 + max_{z\neq s}(min_k(f^s_k-f^z_k)))/ 2 \; \forall z \in Z
\label{eq:fmetrics}
\end{equation}

\subsubsection{Subjective evaluation: architectural preferences}\label{subsec:preferences}

To quantify the \emph{subjective component}, $f_{sub}$, the algorithm makes use of the set of design decisions compiled after each interaction. Taking a candidate solution as a reference, the engineer might highlight a qualitative aspect that he/she considers relevant to appear in a final solution, or focus his/her attention on features to be avoided. Notice that qualitative criteria will mostly be focused on phenotypic aspects of a solution, e.g. whether a software component is meaningful. 
Once the association between the design decision and the architectural preference has been established, the algorithm is responsible for promoting the solutions that satisfy positive preferences and, at the same time, penalizing those presenting undesirable characteristics according to the negative preferences. To do this, $pref_p$ in Eq.~\ref{eq:fpref} measures to what extent an arbitrary solution $s$ satisfies the architectural preference $p$. The returning value, i.e. the \emph{degree of achievement}, lies in the range [0,1] and should be maximized, regardless of whether the engineer's opinion is positive or negative. In addition, each preference has an associated weight, $w_p$, which represents the engineer's confidence in his/her decision. This can be expressed using the Likert-type scale, though the corresponding weight should be scaled in order to ensure that the result remains in the range [0,1]. Here, weights are relative to a unique interaction, so the specific value is computed according to the confidence levels associated to all the evaluations made in the same interaction.

\begin{equation}
f_{sub}(s)=1-1/P\cdot\sum_{p=1}^{P}{(w_p \cdot pref_p(s))} 
\label{eq:fpref}
\end{equation}

\begin{figure*}[!t]
	\centering
	\includegraphics[width=0.5\textwidth]{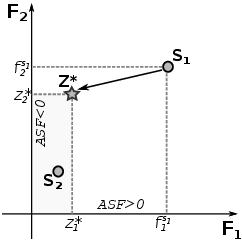}
	\caption{An illustrative example of ASF values}
	\label{fig:asf}
\end{figure*}

At different interactions, the expert will express his/her opinion on either the specific composition of the solutions under evaluation or the values of the returned software metrics. The following list compiles the preference alternatives available for the engineer:

\begin{enumerate}
\item \textit{No preference}. The expert skips providing any opinion.

\item \textit{Best component}. For a given solution, the engineer selects the best component, $c^+$, according to its structure. A preference function $pref_{bc}$ will determine to which extent there are other individuals having a similar component (see Eq. 4). For each component $c$ of a solution, the function $cl()$ extracts its classes, so that the resulting set is compared with the set of classes contained in $c^+$. With this aim, the Jaccard index $J$ is calculated (see Eq. 5). Given a pair of sets, $A$ and $B$, this similarity measure calculates the ratio between the number of common elements and the number of different elements. Finally, the maximum similarity value among the $n$ components comprising the architectural solution is returned as the degree of achievement of this preference.

\begin{eqnarray}
&pref_{bc}=max\{J(cl(c),cl(c^+))\} \ \forall c \in [1,n]\\
\label{eq:bc}
&J(A,B) = |A\cap B| / |A\cup B|\\\nonumber
\label{eq:jaccard}
\end{eqnarray}

\item \textit{Worst component}. In contrast to the previous preference, this preference allows the engineer to express a negative opinion on an observed component $c^-$. Here, the corresponding preference function $pref_{wc}$ penalizes those solutions having a component similar to $c^-$ (see Eq. \ref{eq:wc}).

\begin{equation}
pref_{wc}=max\{1-J(cl(c),cl(c^-))\} \ \forall c \in [1,n]
\label{eq:wc}
\end{equation}

\item \textit{Best provided interface}. The expert may identify an interface $p^+$ of interest for the service interaction specification, even when the component providing it is not properly formed yet. Similar to $pref_{bc}$, the preference function defined by Eq.~\ref{eq:bi} computes the Jaccard index, in this case being used to compare sets of interface operations. The function $op(i)$ serves to extract the operations from an interface $i$, while $in(c)$ compiles the interfaces provided by a component $c$.

\begin{equation}
pref_{bi}=max\{J(op(in(c)), op(p^+))\} \ \forall c\in [1,n]
\label{eq:bi}
\end{equation}

\item \textit{Worst provided interface}. Similar to $pref_{wc}$, it focuses on interfaces instead of components. Eq.~\ref{eq:wi} shows the expression that calculates the preference function $pref_{wi}$, $p^-$ being the rejected interface.

\begin{equation}
pref_{wi}=max\{1-J(op(in(c)), op(p^-))\} \ \forall c\in [1,n]
\label{eq:wi}
\end{equation}

\item \textit{Number of components}. The engineer may be interested in leading the search for solutions with a preferred number of components $n^+$. In this case, $pref_{nc}$ calculates the difference between this number and the current number of components of the given solution, $n$. In Eq.~\ref{eq:nc}, $n_{min}$ and $n_{max}$ are the limits initially set by the engineer to the size of the architecture. An evolutionary consequence, which remains transparent to the expert, is the obvious reduction of the search space.

\begin{eqnarray}
\label{eq:nc}
pref_{nc} =
\begin{cases}
			(n-n_{min})/(n^+-n_{min}) & \text{if $n<n^+$}\\
			1-((n-n^+)/(n_{max}-n)) & \text{if $n\geq n^+$}\\
	\end{cases}
\end{eqnarray}

\item \textit{Metric in a range}. This preference helps the engineer to determine the expected values---or range of values---of a given metric $m$ (see Table~\ref{tab:metrics} in Section~\ref{subsec:metrics}). Having set a maximum ($m_{max}$) and a minimum ($m_{min}$) value for $m$, the preference $pref_{mr}$ penalizes any solution $s$ with a metric value $m^s$ outside this interval. On the contrary, values close to the midrange ($m_{mid}$) are rewarded, as shown in Eq.~\ref{eq:mr}.

\begin{eqnarray}
\label{eq:mr}
&m_{mid} = (m_{max}-m_{min}) / 2\\\nonumber
&pref_{mr} =
\begin{cases}
			0 & \text{if $m^s<m_{min}$}\\
			1-(m^s-m_{mid})/m_{mid} & \text{if $m^s \in [m_{min},m_{max}]$}\\
			0 & \text{if $m^s>m_{max}$}\\
	\end{cases}
\end{eqnarray}

\item \textit{Aspiration levels}. This preference allows the engineer to set the target values for all the metrics. From the evolutionary perspective, aspiration levels~\cite{Miettinen2008} are appropriate to guide the search towards solutions whose objective values are close to the DM expectations. Achievement scalarizing functions (ASFs)~\cite{Miettinen2002} are usually applied to determine to what extent a solution $s$ satisfies the aspiration levels represented in the form of a reference point $z^*$. The ASF here selected, shown in Eq.~\ref{eq:al}, computes a weighted distance in each search direction $k$ so that the overall preference $pref_{al}$ promotes those solutions with a small ASF value. Notice that solutions having better objective values, $f^s_k$, than $z^*$ obtain the maximum degree of achievement for this preference $k$. Assuming equal weights, $w_k$, Fig.~\ref{fig:asf} illustrates both cases. The ASF value for $s_1$ would be determined by the distance in axis $F_1$. ASF is lower than 0 for $s_2$ as it has better objective values than $z^*$. Both the reference point and the weights are provided by the software engineer.

\begin{eqnarray}
\label{eq:al}
&ASF = max\{w_k\cdot(f^s_k-z^*_k)\}\\\nonumber
&pref_{al} =
\begin{cases}
			1 & \text{if $ASF\leq 0$}\\
			1-ASF & \text{if $ASF > 0$}\\
\end{cases}\\\nonumber
\end{eqnarray}

\end{enumerate}

\subsection{Selection, mutation and replacement strategies} \label{subsec:algorithm}

Focusing on the selection mechanism, two parents are selected using binary tournament, one from the population and another from the archive. Since the competition is based on their fitness values, both quantitative and qualitative criteria influence the selection process.

Then, the algorithm applies the mutation operator in order to produce two offspring. 
The mutation operator simulates five different architectural transformations: adding a component ($a$); removing a component ($r$); merging two components ($m$); splitting a component ($s$); and moving a class ($c$)~\cite{Ramirez2015a}. For each individual, the mutator executes a probabilistic roulette with these operations ensuring that any architectural output will be comprised of an allowed number of components, according to the aforementioned thresholds. After selecting an operation and applying it, any mutant not satisfying all the constraints is discarded, and the original solution mutated once again. This process is performed for a maximum of 10 attempts. If the mutator still fails to find a feasible individual, the initial solution is returned. In addition, the operator should be aware of the fact that individuals can contain frozen parts that should not be modified. On the other hand, crossover is not considered because it would hardly generate feasible solutions~\cite{Ramirez2015a}.

Finally, the replacement strategy causes a competition among offspring and current individuals, promoting the survival of those solutions having better fitness values. Solutions discarded by the engineer are progressively removed from the population in order to keep the population size constant. Additionally, if three or more solutions are marked to be removed, two of them would be eliminated in the current generation, while the others will be conveniently penalized in order to ensure their removal in future generations.

\subsection{Archive update mechanism} \label{subsec:archive}

\begin{algorithm}
\caption{Update archive}
\label{alg:archiveUpdating}
\begin{algorithmic}[1]
\REQUIRE{$population$, $archive$}
	\STATE{$archive' \leftarrow archive$}
	\FORALL{$ind \in population$}
		\IF{$ind \notin archive'$}
			\STATE{$accept \leftarrow$false}
			\STATE{$dominated \leftarrow$ solutionsDominatedBy($ind$)}
			\STATE{$r \leftarrow$ preferredRegion($ind$)}
			\STATE{$t \leftarrow$ territorySize($r$)}
			\STATE{$s \leftarrow$ closerSolution($archive'$,$ind$)}
			\STATE{$d \leftarrow$ distance($s$,$ind$)}
			\IF{$ind$ was selected by the user}
				\STATE{$accept \leftarrow$true}
				\IF{$d<t$}
					\STATE{reduceTerritorySize($r$, $t-d$)}
				\ENDIF
			\ELSE 
				\IF{$ind$ is non-dominated}
					\IF{$d>t$}
						\STATE{$accept \leftarrow$true}
					\ELSE
						\STATE{$n \leftarrow $ numberOverlappingTerritories($ind$)}
                        \IF{$n==1$ AND $f_{sub}(ind)>f_{sub}(s)$}
								\STATE{$accept \leftarrow$true}
								\STATE{$archive' \leftarrow (archive' \cap \neg s)$}
						\ENDIF
					\ENDIF
				\ENDIF
			\ENDIF
		\ENDIF
		\IF{$accept$}
			\STATE{$archive' \leftarrow (archive' \cap \neg dominated \cup ind)$}
		\ENDIF
	\ENDFOR
	\RETURN{$archive'$}
\end{algorithmic}
\end{algorithm}

Algorithm~\ref{alg:archiveUpdating} describes the procedure to update the archive, which is partially based on the definition of territories proposed by iTDEA~\cite{Koksalan2010}. However, there are some aspects of the original procedure that have been conveniently adapted in order to apply the proposed fitness function when determining which solutions should be kept in the archive after each generation. Firstly, the method checks that the individual was not added beforehand (line 3). Secondly, it extracts the set of dominated solutions (line 5). Then, the method proceeds like iTDEA (lines 6-9). More specifically, a preferred region for the given individual ($r$) is determined according to a set of weights associated to each objective function~\cite{Koksalan2010}. After finding the archive member $s$ closer to the individual (line 8), the rectilinear distance between them is calculated (line 9). At this point, the decision about whether $ind$ should replace $s$ is not only based on that distance, but also on how much each one satisfies the engineer's preferences. It is a key difference with respect to iTDEA, which would simply discard $ind$ if it lies on the territory associated to $s$. The complete acceptance criteria are defined as follows:

\begin{itemize}
	\item A solution of interest to the engineer is always accepted, regardless of whether it is dominated or not (line 11). In addition, they cannot be removed in subsequent generations. The territory size of the associated region is conveniently reduced when the solution lies on the territory of another solution (lines 12-14).
	
	\item Replacing one non-dominated solution with another belonging to the same region will be permitted if it implies improving $f_{sub}$ (lines 19-26). It should be noted that if the solution overlaps with the territory of two or more archive members (line 20), the action will not be performed in order to avoid discarding an excessive number of solutions.

	\item Dominated solutions are only removed when the individual whose acceptance is being checked is finally added (lines 31-33), as suggested in~\cite{Koksalan2010}.
\end{itemize}

After each interaction, the algorithm reduces the territory size for the region allocating the best solution in terms of $f_{sub}$, allowing a higher density of solutions around.

\subsection{Interaction mechanism}\label{subsec:interaction}

There are two relevant factors that may affect users' fatigue and their loss of interest: the frequency of interaction and the mechanism of selection of solutions. To prevent fatigue, the engineer is able to set the desired number of interaction steps. Then, how these interactions are distributed along the search is automatically determined according to~\cite{Koksalan2010}. Given a maximum number of generations, $g$, $g/3$ generations are executed before the first interaction in order to approximate the PF, while $g/6$ iterations are performed by the algorithm after the last interaction to ensure that human decisions are properly propagated. Between these two points, the user will be able to take action at regular intervals.

A clustering approach is applied in order to select the most representative solutions of the overall population. Notice that it is also important to provide software engineers with information regarding the improvements resulting from their decisions. Therefore, if $m$ solutions are required to be displayed to the expert, a \emph{kMeans++} algorithm~\cite{Arthur2007} selects $m-1$ solutions, looking for diversity with respect to their objective values. The remaining solution is the one with the highest $f_{sub}$ value.

Engineers are also allowed to perform additional actions. Firstly, the most promising solutions can be directly saved to the archive, not only implying that they will be returned at the end of the search process, but also that they could be selected as parents more frequently. Secondly, solutions not satisfying the engineer's expectations could be identified for removal in the replacement phase. Finally, parts of an individual genotype could be frozen as a way to facilitate their appearance in other solutions.

\section{Experimental framework}\label{sec:experimentation}

The IEC model has been coded in Java using JCLEC-MOEA~\cite{Ramirez2015c}. In addition, some supporting libraries have been used to process data\footnote{\url{http://www.uco.es/grupos/kdis/datapro4j}}, extract analysis information from XMI files\footnote{\url{http://www.sdmetrics.com/OpenCore.html}} and implement the clustering procedure\footnote{\url{http://commons.apache.org/proper/commons-math/}}. These algorithm executions aimed at assessing the performance of the evolutionary search were run on an Ubuntu 16.4 computer with 8 cores Intel Core i7 2.67-GHz and 7.79-GB RAM.

The rest of this section explains the empirical methodology conducted to properly respond to research questions \emph{RQ1} and \emph{RQ2}. The parameter set-up and problem instances are detailed next.

\subsection{Methodology}\label{subsec:methodology}

The performance of a new MOEA is usually assessed in terms of quality indicators. Nevertheless, interactive approaches also imply putting the human in the loop in order to perform their evaluation. Both ways are complementary, though the latter requires conducting some particular experimentation and analysis.

Firstly, as posed in \emph{RQ2}, the performance of the multi-objective evolutionary approach has to be proved before the human getting involved. With this aim, a parameter study is required to analyze the behavior of the algorithm regarding the returned number of solutions and the expected trade-off between their quality and diversity. These properties will be evaluated using two quality indicators, hypervolume ($HV$) and spacing ($S$)~\cite{Coello2007}. In addition, the evolutionary performance is compared against the well-known NSGA-II algorithm~\cite{Deb2002}. In both cases, algorithms will be executed 30 times with different random seeds over all the available problem instances.

The significance of the outcomes~\cite{Arcuri2014} is assessed by applying non-parametric statistical tests. More specifically, the Wilcoxon Signed-Rank test will be executed to perform pairwise comparisons, while the Aligned Friedman test will be used when the experiment involves more than two algorithms. An effect size measurement is also considered to assess the performance gain, the Cliff's Delta test being selected to this end. For all experiments, the null hypothesis, $H_0$, establishes that the corresponding algorithms perform equally well at a specific level of significance, 1-$\alpha$.

Once the non-interactive algorithm (hereinafter referred to as bMOEA, \emph{base MOEA}) is properly analyzed and tuned, the interactive approach (iMOEA) will be empirically assessed through the participation of 9 people. All the participants are either students or professionals in the field of computer science, with previous background in software development for a period of 2 to 17 years. More specifically, the experiment is conducted by 1 undergraduate student, 2 master students, 4 software engineers (postgraduates) and 2 academics (PhDs), who face a specific real-world problem instance named Datapro4j (see further details in Section~\ref{subsec:setup}). This software system has been selected as case study because its analysis model could be understandable by the engineer for the time required to conclude the experiment. Additionally, about 33\% of the participants already had some previous development experience with Datapro4j in one form or another.

Interactive sessions have been planned as follows. At the beginning, participants are instructed in the purpose and content of the experiment, as well as in the use of the interactive tool, including the architectural preferences and actions. Given that users are able to visualize the entire architectural model, special considerations are taken to reduce the inherent cognitive burden. On the one hand, internal information about classes is not shown as it remains unaltered from one solution to another, and a printed copy of the input class diagram is also provided during the session. On the other hand, the number of components is restricted according to the configuration parameters, while the mechanism proposed to derive interfaces ensures that the number of interfaces per component is not excessive.

Each participant executes a single run with the best configuration obtained after the parameter study and a different random seed. The interaction scheme consists of 3 stops for the user to evaluate 3 different solutions each time. This way all the participants evaluate the same number of solutions and every interactive execution will provide intermediate results under the same conditions.

Exhaustive execution logs are generated to properly study the behavior of participants and receive relevant feedback from their experience. During the experiment, participants are requested to write down the reasons of their decisions. Similarly, at the very end of the session, they should fill in a form with additional questions about their experience and impressions, as well as free comments or suggestions.

The two planned experiments, i.e. the parameter study and the empirical investigation, serve to respond \emph{RQ2}. On the one hand, the analysis in terms of quality indicators proves the competitiveness of the algorithm from an evolutionary point of view. On the other hand, the usefulness and intuitiveness of the approach can be derived from the information gathered from the interactive session. Furthermore, the influence of the participants' opinion with respect to the sort of the solutions found and the level of metric optimization is assessed by comparing the outcomes of bMOEA and iMOEA. Notice that, in this problem, any configuration is a solution, with the only exception of unfeasible individuals. Therefore, the fitness function is intended to simulate the software engineer's behavior in terms of his/her design preferences, which could not be formulated beforehand, as they are expected to be indicated as the search progresses. Consequently, the solutions returned do not necessarily have to be Pareto optimal, but a reflection close to the expectations of the expert.

\subsection{Algorithm set-up}\label{subsec:setup}

Table~\ref{tab:parameters} shows the list of parameters and their respective values. Overall parameters like the population size and the stopping criteria, as well as those being specific to the problem under study, have been set according to our previous studies~\cite{Ramirez2015a,Ramirez2015b}. Same weights were applied to the objective and subjective evaluations. Furthermore, the influence of the territory size ($\tau$) on the update mechanism of the archive deserves special attention, since it might affect the number of resulting solutions. In this case, three different initial values of $\tau$ are tested during the non-interactive experiment, while recommendations from the authors of iTDEA are followed regarding the mechanism to update it after each interaction~\cite{Koksalan2010}. 
As mentioned in Section~\ref{subsec:methodology}, the interaction scheme ensures that all users interact 3 times with the algorithm, sequentially showing 3 different solutions each time.

\begin{table}[!t]
\begin{center}
\caption{Required parameters and their values}
\begin{tabular}{|l|c|}
\hline
\textit{Parameter}&\textit{Value}\\\hline
Population size & 150\\
Maximum number of evaluations & 24000\\
Minimum number of components & 2\\
Maximum number of components & 6\\
ERP metric weights ($w_{as}$, $w_{ag}$, $w_{co}$, $w_{ge}$) & 1, 2, 3, 5\\
Mutation weights ($w_{a}$, $w_{r}$, $w_{m}$, $w_{s}$, $w_{c}$) & 0.2, 0.1, 0.1, 0.3, 0.3\\
\hline
Fitness weights ($w_{obj}$, $w_{sub}$)  & 0.5, 0.5\\
Initial territory size ($\tau_{0}$) & 0.01, 0.05, 0.1\\
Final territory size ($\tau_{H}$) & 0.005\\
Decreasing factor ($\lambda$) & 0.5\\
Number of interactions & 3\\
Number of solutions per interaction & 3\\
\hline
\end{tabular}
\label{tab:parameters}
\end{center}
\end{table}

Table~\ref{tab:instances} shows the quantitative characteristics of each problem instance. They provide a wide spectrum of complexity regarding the number of classes and relationships, which are classified into associations (\textit{as}), aggregations (\textit{ag}), compositions (\textit{co}), generalizations (\textit{ge}) and dependencies (\textit{de}). The last column indicates the number of candidate interfaces, i.e. the number of relationships whose navigability has been explicitly specified. With the exception of Aqualush\footnote{http://www.ifi.uzh.ch/en/rerg/research/aqualush.html}, a benchmark used for educational purposes, all the problem instances represent working software systems\footnote{http://www.java-source.net/}.

\begin{table}[!t]
\begin{center}
\caption{Problem instances and their characteristics}
\begin{tabular}{|l|c|ccccc|c|}
\hline
\multirow{2}{*}{\textit{Problem}}&\multirow{2}{*}{\textit{Classes}}&\multicolumn{5}{|c|}{\textit{Relationships}}&\multirow{2}{*}{\textit{Interfaces}}\\
			& 		&\textit{as}	&\textit{ag}	&\textit{co}	&\textit{ge}	&\textit{de}	&\\
\hline
Aqualush	&58		&69				&0				&0				&20				&6				&74\\
Datapro4j	&59		&3				&3				&2				&49				&4				&12\\
Java2HTML	&53		&20				&15				&0				&15				&66				&170\\
JSapar		&46		&7				&21				&9				&19				&33				&80\\
Marvin		&32		&5				&22				&5				&8				&11				&28\\
NekoHTML	&47		&6				&15				&18				&17				&17				&46\\
\hline
\end{tabular}
\label{tab:instances}
\end{center}
\end{table}

\section{Analysis of results}\label{sec:results}

\subsection{Evolutionary performance}\label{subsec:performance}

The number of solutions returned to the expert is a key aspect when dealing with real-world decision scenarios and interactive approaches. In the proposed bMOEA, the archive size can be controlled by the parameter $\tau$. Even though the final number of solutions might depend on the way the engineer interacts with the algorithm, providing some guidance to the algorithm in this regard could help.

As explained in Section~\ref{subsec:setup}, the algorithm has been run considering three possible values of $\tau$: 0.01, 0.05 and 0.1. Notice that there is no interaction here, so $\tau$ remains constant along the search. Consequently, it only imposes a restriction with respect to the minimum number of solutions to be returned. Table~\ref{tab:onvg} shows the average archive size for the three aforementioned configurations, including the standard deviation. As might be expected, increasing the value of $\tau$ allows reducing the archive size. Furthermore, the specific problem instance might be another determinant factor.

\begin{table}[!t]
\begin{center}
\caption{Number of solutions of the final archive}
\begin{tabular}{|l|c|c|c|}
\hline
			& $\tau=0.01$		& $\tau=0.05$		& $\tau=0.1$\\
\hline
Aqualush	& $23.77\pm 6.28$ 	& $14.80\pm 2.52$ 	& $9.23\pm 1.87$\\\hline
Datapro4j	& $12.57\pm 3.62$ 	& $8.37\pm 2.06$ 	& $5.73\pm 1.12$\\\hline
Java2HTML	& $140.37\pm 3.03$	& $49.75\pm 5.17$ 	& $17.48\pm 1.92$\\\hline
JSapar		& $21.67\pm 4.99$ 	& $12.20\pm 2.94$ 	& $8.00\pm 1.41$\\\hline
Marvin		& $13.27\pm 4.02$ 	& $8.27\pm 2.02$ 	& $5.93\pm 1.29$\\\hline
NekoHTML	& $20.07\pm 4.56$ 	& $11.60\pm 3.20$ 	& $7.97\pm 1.33$\\
\hline
\end{tabular}
\label{tab:onvg}
\end{center}
\end{table}

Given that a limited number of solutions could compromise the expected trade-off between convergence and diversity, a further analysis of the quality of the solutions is carried out in terms of $HV$ and $S$. Firstly, the spacing values for the three configurations are studied in order to confirm that the use of territories allows our bMOEA to effectively preserve diversity while maintaining an affordable number of alternatives to be returned to the participant.

Table~\ref{tab:s} shows the obtained results for each problem instance, where higher values are better. As can be seen, setting too low values of $\tau$ might demote the expected diversity, mainly because of the increase of the number of solutions returned by the algorithm. Rankings reported by the Aligned Friedman test are also included in Table~\ref{tab:s}. To confirm the existence of statistical differences using this test, the statistic $z$--distributed according to a chi-square distribution with 2 degrees of freedom--should be compared to a critical value of that distribution. Given that the obtained statistic, $z=4.1477$, is smaller than the critical value (5.9915) for $\alpha=0.05$, $H_0$ cannot be rejected, that is, bMOEA behaves similarly in terms of spacing for the three proposed configurations.

\begin{table}[!t]
\begin{center}
\caption{Results for Spacing ($S$)}
\begin{tabular}{|l|c|c|c|}
\hline
			& $\tau=0.01$		& $\tau=0.05$		& $\tau=0.1$\\
\hline
Aqualush	& $0.039\pm 0.023$ 	& $0.045\pm 0.028$ 	& $0.048\pm 0.020$\\\hline
Datapro4j	& $0.038\pm 0.034$ 	& $0.044\pm 0.045$ 	& $0.050\pm 0.027$\\\hline
Java2HTML	& $0.021\pm 0.002$ 	& $0.022\pm 0.004$ 	& $0.030\pm 0.009$ \\\hline
JSapar		& $0.041\pm 0.021$ 	& $0.064\pm 0.017$ 	& $0.063\pm 0.031$\\\hline
Marvin		& $0.031\pm 0.009$ 	& $0.029\pm 0.015$ 	& $0.035\pm 0.019$\\\hline
NekoHTML	& $0.033\pm 0.011$ 	& $0.032\pm 0.017$ 	& $0.037\pm 0.019$ \\\hline
\textbf{Ranking} & 14.500 & 9.500 & 4.500 \\\hline
\end{tabular}
\label{tab:s}
\end{center}
\end{table}

In contrast, as can be observed in Table~\ref{tab:hv}, when comparing the three configurations in terms of $HV$, better values are now obtained for $\tau=0.01$, though the Aligned Friedman test reveals that there are not statistical differences ($z=4.0730$) at a confidence level (CL) of 95\%. Consequently, $\tau=0.05$ is chosen as the most appropriate value for our algorithm, since it achieves the best trade-off between both indicators and the difference with the corresponding best possible configuration is not statistically significant.

\begin{table}[!t]
\begin{center}
\caption{Results for hypervolume ($HV$)}
\begin{tabular}{|l|c|c|c|}
\hline
			& $\tau=0.01$		& $\tau=0.05$		& $\tau=0.1$\\
\hline
Aqualush	& $0.637\pm 0.012$ 	& $0.620\pm 0.015$ 	& $0.599\pm 0.017$\\\hline
Datapro4j	& $0.658\pm 0.017$ 	& $0.639\pm 0.017$ 	& $0.614\pm 0.022$\\\hline
Java2HTML	& $0.282\pm 0.026$ 	& $0.278\pm 0.027$ 	& $0.261\pm 0.035$ \\\hline
JSapar		& $0.555\pm 0.015$ 	& $0.552\pm 0.018$ 	& $0.530\pm 0.018$\\\hline
Marvin		& $0.616\pm 0.008$ 	& $0.614\pm 0.007$ 	& $0.604\pm 0.012$\\\hline
NekoHTML	& $0.602\pm 0.012$ 	& $0.586\pm 0.015$ 	& $0.567\pm 0.015$\\\hline
\textbf{Ranking} & 3.667 & 9.333 & 15.500 \\\hline
\end{tabular}
\label{tab:hv}
\end{center}
\end{table}

In addition to the influence of $\tau$, it would be also interesting to find out to what extent the proposed algorithm provides a similar performance that those MOEAs aimed at returning an approximation of the whole PF. In fact, NSGA-II has also been considered here due to its ability to effectively guide the search towards non-dominated solutions and the lack of an explicit mechanism to limit the number of final solutions. Table~\ref{tab:nsga2} shows the results for $S$ and $HV$, as well as the number of non-dominated solutions found. Although NSGA-II provides values of $HV$ slightly higher than our algorithm (see Table~\ref{tab:hv} for $\tau=0.05$), it is worth noticing the difference with respect to the spacing indicator. Besides, the high number of returned solutions could complicate the decision-making process.

Pairwise comparison is performed here in order to precisely compare their performance. Regarding $HV$, the Wilcoxon test reveals that NSGA-II performs better than our algorithm with a CL of 90\% ($p-value=0.0938$), even though this difference is classified as \textit{negligible} by the Cliff's Delta test. On the contrary, our proposal is statistically better than NSGA-II in terms of $S$ with a CL of 95\%, $p-value=0.0313$, the difference between both algorithms being classified as \textit{large}. In this sense, our algorithm is a competitive alternative against NSGA-II, since it is able to find not only high quality but also representative solutions, even when the archive size has to be limited.

\begin{table}[!t]
\begin{center}
\caption{Results obtained by NSGA-II}
\begin{tabular}{|l|c|c|c|}
\hline
			&\textit{Num. of solutions}&\textit{Spacing}&\textit{Hypervolume}\\
\hline
Aqualush	& $147.73\pm 11.84$ & $0.018\pm 0.008$	&$0.635\pm 0.015$ \\\hline
Datapro4j	& $148.67\pm 4.25$ 	& $0.013\pm 0.013$	&$0.645\pm 0.010$ \\\hline
Java2HTML	& $148.87\pm 2.96$ 	& $0.009\pm 0.008$	&$0.404\pm 0.048$ \\\hline
JSapar		& $150.00\pm 0.00$ 	& $0.018\pm 0.007$	&$0.547\pm 0.014$ \\\hline
Marvin		& $140.60\pm 23.22$ & $0.014\pm 0.005$	&$0.618\pm 0.009$ \\\hline
NekoHTML	& $150.00\pm 0.00$ 	& $0.016\pm 0.007$	&$0.596\pm 0.013$ \\\hline
\end{tabular}
\label{tab:nsga2}
\end{center}
\end{table}

\subsection{Use of interactive mechanisms}\label{subsec:useprefs}

\begin{table}[!t]
\begin{center}
\caption{Architectural preferences applied during the experiment}
\begin{tabular}{|l|c|c|c|}
\hline
\textit{Architectural Preference}&\textit{\% Selected}&\textit{Usefulness} & \textit{Intuitiveness}\\
\hline
No preference				& 22.22\%	& 6.44	& 7.33\\\hline
Best component				& 29.63\%	& 7.44	& 7.44\\\hline
Worst component				& 23.46\%	& 7.22	& 7.33\\\hline
Best provided interface		& 2.47\%	& 5.29	& 6.38\\\hline
Worst provided interface	& 0.00\%	& 4.71	& 6.38\\\hline
Number of components		& 17.28\%	& 7.50	& 7.33\\\hline
Metric in range				& 2.47\%	& 4.17	& 5.44\\\hline
Aspiration levels			& 2.47\%	& 5.80	& 5.22\\\hline
\end{tabular}
\label{tab:preferences}
\end{center}
\end{table}

\begin{figure*}[!t]
	\centering
	\includegraphics[width=0.9\textwidth]{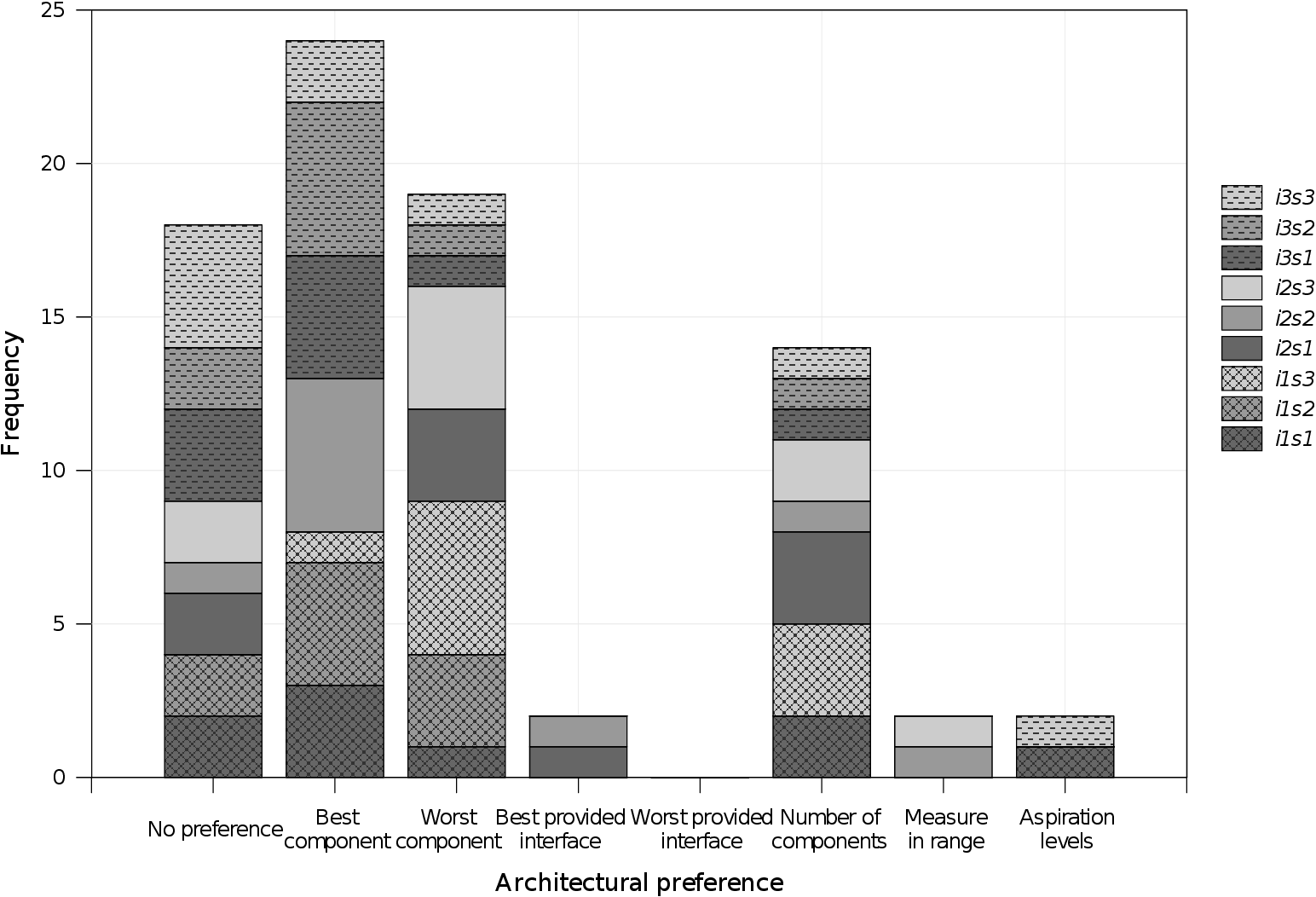}
	\caption{Frequency and moment of selection of each architectural preference}
	\label{fig:frequency}
\end{figure*}

A key aspect of the proposed interactive approach is that the engineer is able to evaluate the solutions provided by the algorithm in qualitative terms, e.g. by indicating both positive and negative preferences that might influence the subsequent search process. It is interesting to observe how these architectural preferences are selected, and how useful and intuitive participants consider their application. These two factors are scored by users on a scale between 1 (\textit{poor}) and 8 (\textit{excellent}). Table~\ref{tab:preferences} shows the frequency of use of preferences, and the average rating for usefulness and intuitiveness.

Within the preferences group, participants generally focus their attention on the internal structure of the components. In fact, indicating the preferred number of components that should comprise the architectural specification is also a frequent choice. In contrast, preferences related to interfaces have been rarely selected. It is likely that users consider a priority to find a proper structure at first, and they omit any further detail on the component interaction, even when it could be a factor to refine the search by filling components with the most appropriate interacting constituents, i.e. classes and relationships. Participants also avoid setting specific values to software metrics, possibly because they consider this could not lead to such a straightforward and tangible result. Finally, it is also a common practice not to indicate any architectural preference. Some participants pointed out that they just wanted to observe how the algorithm could evolve by itself, whilst others found it a convenient way to deal with uncertainty about making a precise judgment. In any case, the applicability of architectural preferences seems to be related to their intuitiveness and usefulness, according to users' scores.

In order to examine the behavior of the participants, it is also interesting to study whether their design decisions are somehow related to the interaction moment. Fig.~\ref{fig:frequency} shows the total number of occurrences of the different preferences at each specific interaction point, where $i$ stands for the number of interaction break, and $s$ for the solution position in the group of three. Notice that, apart from the omission of choice, the three most intuitive and useful preferences---according to the user rating---were applied throughout the entire search process. Nevertheless, some additional patterns can be still observed. For instance, during the initial interactions, users tend to express negative opinions (e.g. \textit{worst component}) or to indicate some overall restriction (e.g. \textit{number of components}) in order to reach an expected solution. However, as the evolution progresses, positive opinions become more frequent because better solutions are returned.

\begin{table}[!t]
\begin{center}
\caption{Other actions taken during the interactive session}
\begin{tabular}{|l|c|c|c|}
\hline
\textit{Optional action}&\textit{\% Selected}&\textit{Usefulness}&\textit{Intuitiveness}\\
\hline
Add to archive				& 34.78\%	& 6.11	& 6.89 \\\hline
Remove from population		& 30.43\%	& 5.89	& 6.89 \\\hline
Freeze components			& 34.78\%	& 7.44	& 7.44 \\\hline
Stop search					& 0.00\%	& 5.14	& 7.44 \\\hline
\end{tabular}
\label{tab:actions}
\end{center}
\end{table}

Apart from indicating a preference, participants have the opportunity to take additional actions, such as adding the solution to the archive, definitely removing it, or even freezing a specific part of an architecture. They could also stop the search process to get the current archive. It is noticeable how these actions are not frequently selected by the user. Actually, 66\% of the participants applied at least one of these actions but only once or twice. Table~\ref{tab:actions} summarizes how many times each action was taken, and how users rated their usefulness and intuitiveness. Participants never stopped the search, as they did not even find it so useful, probably because of the limits in the number of iterations and time constraints. When applied, adding solutions and freezing components served the participants to reinforce their preferences, which also allowed them to perceive the effect of these actions in subsequent interactions.


\subsection{Impact of subjective evaluation}\label{subsec:impact}

Without the user interaction, the proposed algorithm would be guided by objective measures like other non-interactive approaches. The participation of the user might apparently distort the solutions returned from a merely evolutionary perspective. Nevertheless, notice that the real power of the iMOEA lies in its ability to reinforce certain solutions that engineers might find close to their expectations. This is an important aspect to be considered in domains like search-based software design, where a number of objective measures is not enough to evaluate the know-how, previous experiences and overall expectations of the user. Next, the actual influence of the subjective decisions made along the interactive process in the quality of the solutions is discussed.

\begin{table*}[!t]
\begin{center}
\caption{Evolution of software metrics}
\scalebox{0.65}{
\begin{tabular}{|l|ccc|ccc|}
\hline
\multirow{2}{*}{}	&\multicolumn{3}{|c|}{bMOEA}	&\multicolumn{3}{|c|}{iMOEA}\\\cline{2-7}
					&ICD		&ERP		&GCR		&ICD		&ERP		&GCR\\
\hline
Initial population	&$0.619\pm0.009$	&$0.716\pm0.009$	&$0.689\pm0.010$	&$0.548\pm0.006$	&$0.712\pm0.006$	&$0.685\pm0.006$\\
Final population	&$0.640\pm0.035$	&$0.030\pm0.013$	&$0.027\pm0.012$	&$0.366\pm0.066$	&$0.168\pm0.084$	&$0.165\pm0.085$\\
\hline
Initial archive		&$0.476\pm0.021$	&$0.518\pm0.053$	&$0.491\pm0.053$	&$0.481\pm0.036$	&$0.520\pm0.074$	&$0.483\pm0.070$\\
Final archive		&$0.419\pm0.029$	&$0.133\pm0.035$	&$0.121\pm0.033$	&$0.414\pm0.041$	&$0.147\pm0.046$	&$0.139\pm0.042$\\
\hline
\end{tabular}
}
\label{tab:evol-metrics}
\end{center}
\end{table*}

According to the results from Section~\ref{subsec:performance}, bMOEA and iMOEA behave similarly from an evolutionary perspective. For instance, for iMOEA, the average value of $HV$ is equal to 0.6368, which is quite close to the result obtained by bMOEA, $HV=0.6391$ (see Table~\ref{tab:hv} for the Datapro4j instance). Even two participants reached higher values for $HV$ than bMOEA. In terms of $S$, for iMOEA the 9 values lie on the range [0.0143, 0.0859], the average value ($S=0.0433$) being a little less than that obtained by bMOEA ($S=0.0439$). Here, the observed difference in iMOEA is mainly due to the activation of the archive update mechanism, which would be reacting to completely different user actions. In fact, the average number of solutions stored in this set increased from 8.37 to 11.56, returning archives with up to 19 solutions for iMOEA.

Bringing the human in the loop influences the reached trade-off among software metrics. According to the conducted experimentation, most of the design decisions made by the participants were implicitly directed towards the increase of the ICD metric, which otherwise would tend to be demoted by the evolutionary algorithm in favor of ERP and GCR. More specifically, Table~\ref{tab:evol-metrics} shows the average value for each metric in both the initial and the final population. These values are also obtained for the solutions belonging to the archive. It should be noted that all values belong to the range [0,1], 0 being the optimum. For bMOEA, ERP and GCR are greatly improved in the regular population, whereas ICD remains quite constant or even increases. On the contrary, iMOEA is able to reduce ICD significantly without causing a dramatic increase of the other metrics with respect to the values reached by bMOEA. Focusing on the archive, these differences are not so evident. Notice that the archive only stores non-dominated solutions well distributed over the PF, the obtained average values representing better trade-offs among the three metrics. In this sense, it is likely that interesting solutions from the engineer's point of view, e.g. those with low ICD values, represent dominated solutions. This explains why ICD values are higher in the archive than in the regular population. In this case, these solutions cannot appear in the archive unless the user explicitly indicate that they should be added.

\begin{figure}[!t]
	\centering
	\includegraphics[width=0.95\textwidth]{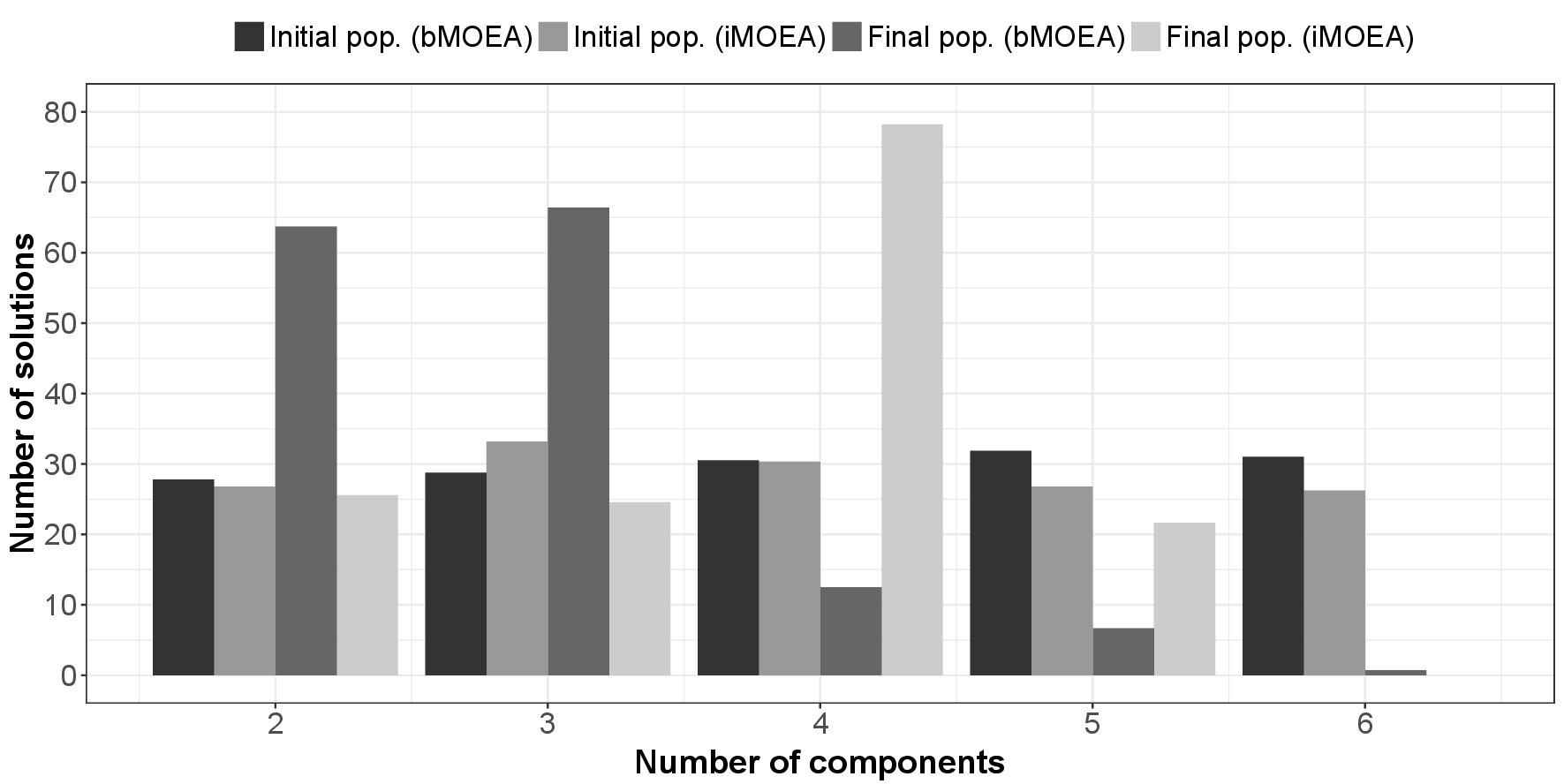}
	\caption{Variation in the size of the solutions during the evolutionary process}
	\label{fig:distribution}
\end{figure}

With iMOEA, software engineers may also recommend their preferred structure for the architectural solutions. As discussed in Section~\ref{subsec:useprefs}, most of the participants selected the architectural preference \emph{number of components} at some point in the interactive session. More precisely, many of them agreed that 4 components was an appropriate value for this problem instance. As can be seen in Fig.~\ref{fig:distribution}, this choice drastically affected the evolution, and iMOEA rapidly discarded solutions of other sizes. In fact, notice that, even when both bMOEA and iMOEA start with a similar distribution of solutions regarding their number of components, bMOEA finally leads the search towards architectural solutions having 2 or 3 components. It is caused by the optimization of the objective criteria, since ERP and GCR can be reduced by creating large components.

\subsection{Comparison between solutions}

A complementary view of the outcomes of the interactive session can be made focusing on the type of solutions found, and their similarity with the manually produced architecture. In addition, it also serves to analyze to what extent participants' decisions might differ from those made by a search process only guided by design metrics. As a reference, the architectural specification of the case study, \emph{Datapro4j}, was originally comprised of 4 components: \emph{Datasets}, \emph{Columns}, \emph{Algorithms (a.k.a. Strategies)} and \emph{Datatypes}. Please notice that this comparison is made against the original source, human-designed specification, which should not strictly comply with the design requirements implied by the design metrics, as they could have consider others as well.

As a result of the interactive session, 104 solutions were returned as part of the 9 final archives. From this set, 6 solutions (5.77\%) contain a component that is equal to one of those specified by the original designers. These solutions were obtained from 5 different participants (55.56\% of the executions), but only one of them explicitly stored the solution in the archive. If approximations to the source architecture are taken into account\footnote{Here, a component having all the classes of the original component with a margin of error of $\pm$2 classes has been considered as a valid approximation.}, the percentage of solutions would increase to 76,92\%. In this scenario, all the participants were able to find at least two alternative solutions with these characteristics. In addition, 3 out of 4 original components were approximated at least once, \emph{Algorithms} being the most frequently identified component (in 52 of the 104 solutions). It is worth mentioning that 7 solutions archived by the participants contained that component, which was also frozen in 4 of them. This suggests that the algorithm can achieve 'human-looking' solutions with the assistance of the user, requiring just a few manual modifications to reproduce the original architecture.

After conducting a similar analysis on the non-interactive algorithm, it can be observed that an exact reproduction of at least one original component is found in 18 of the 251 solutions (7.17\%). However, they all are generated by only 8 of the 30 executions (26.67\%). When approximate components are considered, percentages increase up to 73.31\% and 100\%, respectively. In this case, the algorithm provides good approximations of 3 components, two of them being the same that those identified by iMOEA. The components appearing more often are those comprised of less classes and presenting less interactions, i.e. \emph{Dataypes} and \emph{Algorithms}, as they can be more easily isolated. Therefore, the interaction with the user could also help the algorithm to find a good separation of the most coupled functionalities.

It is worth mentioning that even when the algorithm is able to find similar solutions to those specified manually, the interactive approach takes advantage of human design abilities to identify core functionalities and thus is able to produce meaningful components. In addition, mechanisms like freezing parts of the solutions and their external storage may reinforce the search in order to rapidly propagate human design decisions. In contrast, the influence of the stochastic nature of the evolutionary process can be mitigated.

\subsection{Human experience}\label{subsec:human}

\begin{figure}[!t]
	\centering
	\includegraphics[width=0.8\textwidth]{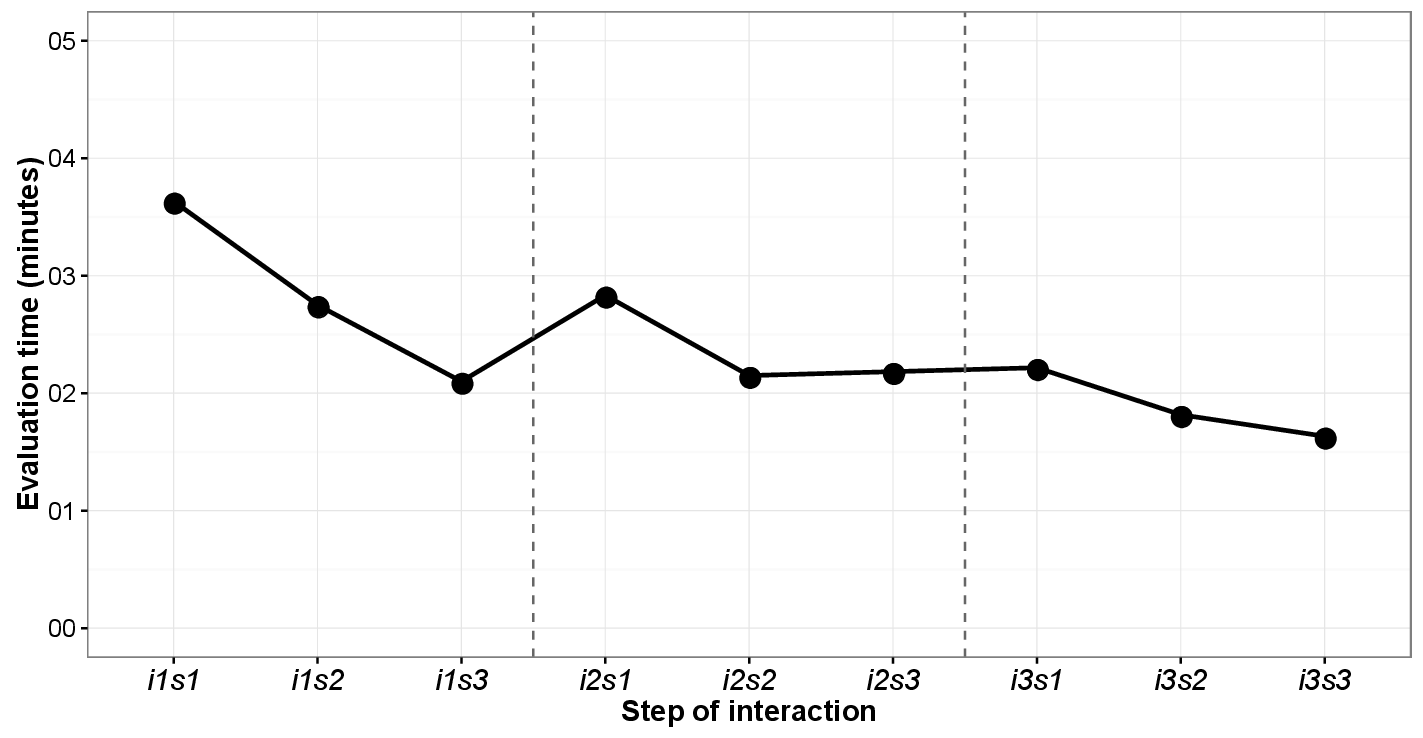}
	\caption{Average evaluation time during interactions}
	\label{fig:evaltime}
\end{figure}

In the context of software design, analyzing a number of full architectural models requires a major effort for the expert. Therefore, steps should be taken in order to alleviate the burden of successive evaluations, such as restricting the number of interactions with the algorithm, or reducing the spent time. With this aim, participants were asked to respond to a survey---scores between 1 (\textit{completely disagree}) and 8 (\textit{completely agree})---related to their fatigue at the end of their interaction. Their replies indicate that, even though they partially disagree with the idea that the interactive session took too long (3.44), they mostly recognized that they were paying more attention at the beginning of the process (5.13). This is clearly reflected by the average time spent per evaluation (see Fig.~\ref{fig:evaltime}). Notice that users take more time to evaluate solutions shown during the initial interaction, probably because of a greater interest and less knowledge on the problem and the process itself. However, as the search process advances and they acquire more knowledge about candidate solutions, users tend to reduce their degree of interest in exploring new alternatives from the returned solutions, while improving their ability to process the information displayed by the interactive tool. This effect is also noticeable for the first solution shown in every interaction. 

As for their feedback, users pointed out that they found promising the idea of having tools to support design tasks (6.89), since they consider that it helps reducing the effort that the design process implies (6.50). The overall perception was that the algorithm could provide interesting solutions (6.13), and it even helped them to discover new design alternatives that they had not thought about (6.38). Finally, participants also made diverse suggestions:

\begin{itemize}
	\item To extend the number of architectural preferences and actions available, as well as to allow applying more than one preference to the same solution.
    \item To allow the user to directly manipulate the provided solution, e.g. by splitting components or moving elements.
    \item To enable participants to undo previous actions and reconsider their decisions, as well as to allow exploring the candidate solutions at once, instead of sequentially.
\end{itemize}

\section{Threats to validity}\label{sec:threats}

The experimental and empirical nature of this study places certain limitations, which are discussed next in terms of validity threats.

Internal validity refers to those aspects of the experimentation that cannot ensure the causality between the hypothesis and the obtained results. In this regard, comparisons between algorithms are based on 30 independent runs in order to deal with the intrinsic randomness of evolutionary algorithms. Appropriate statistical tests have been applied to draw conclusions about the performance of the algorithms in terms of two commonly used quality indicators. As for the experiment putting the human in the loop, the relative small sample size would represent the main threat to internal validity. Nevertheless, 9 or even smaller sizes are common and properly accepted for interactive studies in SBSE~\cite{Simons2014,Marculescu2015}, since the motivation behind this kind of experimentation is mostly focused on the analysis of the human experience, the usefulness of the approach, and the contribution of human-made decisions to the evolutionary process.

The design of the interactive experiment poses additional threats to construct validity. Focusing on the selection of participants, none of them had previous background on the use of interactive algorithms, though some of them were familiar with evolutionary computation in domains different from SBSE. On the contrary, a few participants had some prior knowledge about the system under study, which was intentionally selected in order to reflect the diversity of the practical scenario.

The threat caused by the user fatigue was controlled by applying a fixed interaction scheme, where each participant performs only one execution within an interactive session never longer than 1 hour. It also suitable to mitigate the learning effect. Besides, the visualization and evaluation of complete architectural models might produce a heavy cognitive load to the user. The proposed evaluation method tries to overcome this situation by focusing on the qualitative assessment of parts of the solution, and it has been conceived to be open to other complementary mechanisms.

External validity is related to the generalization of the experimental results. Although participants are mostly working in an academic context, some of them have previous experience in the industry or are currently working in an industrial setting. Nevertheless, given that they all are software engineers, the experiment has served to confirm the benefit of the interactive approach as a supporting mechanism to support the SE professional in the understanding of the underlying architecture of a real software system. 

\section{Concluding remarks}\label{sec:conclusion}

This paper presents an interactive multi-objective evolutionary algorithm aimed at supporting software engineers during the early analysis process. The combination of multi-objective optimization techniques with the so-called architectural preferences guides the search towards the joint optimization of both objective and subjective criteria. Both types of evaluation depend on the specific characteristics of the architecture optimization problem to be addressed, even so the adaptation of the proposed algorithm in order to solve other design tasks would only require the redefinition of specific quality criteria, e.g. the software metrics and architectural preferences. Under the assumption that engineers might detect more easily those model elements that they dislike when analyzing complex architectural specifications, they have been also provided with the possibility to indicate negative opinions on candidate solutions. 

As for its validation, the proposed approach has been compared against a well-known multi-objective algorithm like NSGA-II. To study its suitability for bringing the human in the loop, the algorithm has also been validated with a representative number of users with different expertise levels, who have participated in interactive sessions. Results show that the interactive approach is able to manage the expected trade-off between specific requirements of a real decision scenario: good enough solutions, variety of alternatives and a restricted number of solutions. Furthermore, the use of architectural preferences as a mechanism for the subjective, qualitative evaluation helps to overcome the limitations related to the use of numerical ratings, as usually proposed by other IEC approaches.

To conclude, such a human in the loop approach would definitely allow software engineers to actively participate in the generation and evaluation of different design alternatives, providing the search algorithm with accurate information concerning their real expectations and, consequently, leading to more satisfactory results. In the future, we intend to consider the suggestions made by the participants to improve the interactive experience, and analyze whether their abilities to recognize promising parts of a solution might help to improve the search performance. In addition, the proposed evaluation mechanism could be also applied to other multi-objective decision scenarios like the engineering field~\cite{Kundu2015,Marler2004}, where the opinion and knowledge of experts may suppose a significant difference to reach their expectations~\cite{Martinez2007}.

\section*{Acknowledgments}
This work was supported by the Spanish Ministry of Economy and Competitiveness [projects TIN2014-55252-P, TIN2017-83445-P]; the Spanish Ministry of Education under the FPU program [grant FPU13/01466]; and FEDER funds.



\section*{References}


\end{document}